\documentclass[12pt]{article}
\usepackage{amsmath}
\usepackage{graphicx}
\usepackage{enumerate}
\usepackage{natbib}
\usepackage{url} 
\usepackage{microtype}
\usepackage[utf8]{inputenc} 
\usepackage[T1]{fontenc}    
\usepackage{hyperref}       
\usepackage{url}            
\usepackage{booktabs}       
\usepackage{amsfonts}       
\usepackage{nicefrac}       
\usepackage{microtype}      
\usepackage{lipsum}		
\usepackage{graphicx}
\usepackage{natbib}
\usepackage{doi}
\usepackage{bbm}
\usepackage{dirtytalk}
\usepackage{amsmath}
\usepackage{array}
\usepackage{lineno}
\usepackage{booktabs}
\usepackage{tabularx}
\usepackage{lineno}

\newcommand{\blind}{1}

\begin{document}

\bibliographystyle{plainnat}

\def\spacingset#1{\renewcommand{\baselinestretch}%
{#1}\small\normalsize} \spacingset{1}


\if1\blind
{
  \title{\bf Semiparametric Discovery and Estimation of Interaction in Mixed Exposures using Stochastic Interventions}
  \author{David B. McCoy \thanks{Corresponding author: david\_mccoy@berkeley.edu. The authors gratefully acknowledge funding for Core E of the NIEHS Superfund Center at Berkeley funded by NIH grant P42ES004705}\hspace{.2cm}\\
    Division of Biostatistics, University of California, Berkeley, USA\\
    and \\
    Alan Hubbard \\
    Division of Biostatistics, University of California, Berkeley, USA \\ 
    and \\ 
    Mark van der Laan \\ 
    Division of Biostatistics, University of California, Berkeley, USA \\ 
    and \\
    Alejandro Schuler \\ 
    Division of Biostatistics, University of California, Berkeley, USA \\ }
  \maketitle
} \fi

\if0\blind
{
  \bigskip
  \bigskip
  \bigskip
  \begin{center}
\setlength{\baselineskip}{1.5\baselineskip}
    {\LARGE\bf Semiparametric Discovery and Estimation of Interaction in Mixed Exposures using Stochastic Interventions}
\end{center}
  \medskip
} \fi

\bigskip
\begin{abstract}

This study introduces a nonparametric definition of interaction and provides an approach to both interaction discovery and efficient estimation of this parameter. Using stochastic shift interventions and ensemble machine learning, our approach identifies and quantifies interaction effects through a model-independent target parameter, estimated via targeted maximum likelihood and cross-validation. This method contrasts the expected outcomes of joint interventions with those of individual interventions. Validation through simulation and application to the National Institute of Environmental Health Sciences Mixtures Workshop data demonstrate the efficacy of our method in detecting true interaction directions and its consistency in identifying significant impacts of furan exposure on leukocyte telomere length. Our method, called InterXshift, advances the ability to analyze multi-exposure interactions within high-dimensional data, offering significant methodological improvements to understand complex exposure dynamics in health research. We provide peer-reviewed open-source software that employs or proposed methodology in the \texttt{InterXshift} R package.

\end{abstract}

\noindent%
\vfill

\newpage
\spacingset{1.9} 
\section{Introduction}
\label{s:intro}

In environmental health research, understanding the impact of simultaneous multiple exposures on health outcomes is crucial, yet challenging. Traditional epidemiological studies frequently rely on generalized linear models (GLM) to assess main effects, often neglecting complex interactions within exposure mixtures, potentially oversimplifying the intricate dynamics at play \cite{Wang2018, Garcia-Villarino2022}. Although methods such as Principal Component Analysis (PCA) and penalized regressions, such as LASSO and Ridge, address issues of high dimensionality and multicollinearity, they fall short in interpretability, especially in regulatory and policy contexts where specific guidance on exposure reduction is needed \cite{Roberts2006, Wang2018, McEligot2020}. The limitations of existing approaches, including quantile sum g-computation and linear models without interaction terms, risk simplifying the nuanced effects of interactions \cite{Keil2019}. Bayesian Kernel Machine Regression (BKMR) offers a more nuanced framework that employs Bayesian hierarchical models and kernel machine learning to evaluate high-dimensional exposure effects, providing uncertainty quantification through posterior distributions \cite{Bobb2014}. Despite BKMR's advantages, practical challenges emerge, particularly with defining interactions in high-dimensional settings (which two or three-way interactions to examine), which lead to combinatorial complexity and prioritization challenges for researchers. Given that interactions in a mixed-exposure setting are not known a priori, the number of interactions to test leads to multiple testing issues in any statistical method and underscores the necessity for our proposed method. We begin with a model-agonistic definition of interaction, define oracle target parameters for synergy and antagonism, and then employ cross-validated targeted maximum likelihood estimation (CV-TMLE) paired with data-adaptive target parameters  \cite{Hubbard2016, Zheng2010} to both discover the top set of synergistic and antagonistic interactions using g-computation and then provide efficient estimation of the interaction parameter for the discovered subset using TMLE. Although interaction search methods exist based on variance decomposition \cite{Lengerich2020} and forward selection \cite{Narisetty2019Selection} our approach differs by focusing on a causal interaction target parameter and employs data-adaptive target parameters (sample splitting) which uses training folds for interaction screening and validation folds for estimation. 

Important insights have been provided on how to define interactions in a causal inference framework \cite{vanderweele2009distinction, vanderweele2014tutorial, vanderweele2007directed, rothman1980concepts}. Much of this work has done so in the context of binary explanatory variables of interest. However, the wide range of exposures that characterize environmental investigations often render standard binary causal models inadequate. This accentuates the need for a versatile, semiparametric interaction definition that can accommodate a spectrum of data types, a current gap in methodologies. Our proposed approach, termed \texttt{InterXshift}, synthesizes  two critical phases of analysis, discovery and estimation, using the same data. We define interaction as the comparison of a joint stochastic shift intervention \cite{diaz_2012} (say, increasing exposure 1 and exposure 2 simultaneously by a fixed amount) compared to the sum of shifting each exposure by the same amount individually. We screen for interactions in training folds using g-computation to estimate this interaction parameter, evaluating all two-way interactions, and selecting a subset of the top antagonistic and synergistic interactions. Subsequently, we estimate our interaction parameter using targeted learning in validation folds, which provides us with debiased estimates of these subsets of interactions with confidence intervals. Because we define our oracle parameters as the top subset of synergistic/antagonistic interactions, we can pool these estimates across the folds, leverage the full data used for making these estimates. Our target parameter evaluates the effect of feasible stochastic shift interventions on exposure distributions, as conceptualized in the literature \cite{diaz_2012}.  Such interventions are highly relevant to public policy, as they often mirror the realistic impact of regulatory measures that aim to slightly reduce, rather than completely eliminate or uniformly standardize, exposures across a population. 

The structure of this paper is as follows. Section 2.1 provides a background on the semiparametric methodology that underpins stochastic interventions. Sections 2.2 and 2.3 delineate the interaction target parameter and the essential assumptions for our causal interpretations. Section 3. discusses the TMLE estimator. In Section 4. we discuss how to discover interaction through g-computation. Section 5. presents simulation results, underscoring the asymptotic unbiasedness of our estimates. Section 6 demonstrates \texttt{InterXshift}'s applicability, juxtaposing it against existing techniques and its relevance in real-world scenarios using synthetic data from the National Institute of Environmental Health Sciences (NIEHS). We provide real-world results when applying \texttt{InterXshift} to National Heath and Nutrition Examination Survey, where we investigate the effects of 18 POPs on telomere length. In Section 7 we wrap up with a comprehensive discussion of strengths and weaknesses and provide our software.

\section{Data and Parameter of Interest}
\subsection{Overview}
At a high level, consider that an analyst has data on a mixture of 15 chemicals (e.g. 15 Persistent Organic Pollutants (POPs)). Our approach first splits the entire data into discovery and estimation samples. In the discovery sample, we employ ensemble machine learning to identify interactions through g-computation of our predefined interaction target parameter. For example, consider that we find an interaction between a POP-1 and POP-6 through g-computation. We then use the estimation portion of the data to estimate our interaction target parameter using targeted maximum likelihood estimation. As discussed, our interaction parameter is based on comparing a joint stochastic shift, that is, shifting both exposure simultaneously and observing the expected outcome and comparing this to the summed expected outcomes given individual shifts of each POP independently. We cycle through the data, for example, 10 times, and average estimates for the interaction ranks across the 10 folds using cross-validated targeted maximum likelihood estimation (CV-TMLE). Because our estimand depends on the data, if the full data were used to both discover and estimate, our statistical inference suffers from overfitting; thus, this data-splitting is necessary. With this example in mind, subsequent sections will delve deeper, elucidating each component of our parameter for enhanced clarity.

\subsection{Notation and Framework}
Building upon the concepts presented in \cite{diaz_2012} on stochastic interventions, we generalize this work to a target parameter that compares the expected outcome under a joint shift (here we focus on two exposures for brevity). As such, we first describe our target parameter for a stochastic shift of multiple exposures (which deviates little from estimating the impact of a stochastic shift of an individual variable) and then show how we can build our target parameter for interactions from this joint-shift target parameter. Consider an experiment in which a vector of exposure variables \(\mathbf{A}\), a continuous or binary outcome \(Y\), and a set of covariates \(W\) are measured for \(n\) randomly sampled subjects. Let \(O=(W, \mathbf{A}, Y)\) represent a random variable with distribution \(P_0\), and \(O_1,...,O_n\) represent \(n\) i.i.d. observations of \(O\).  The true distribution \(P_0\) of \(O\) can be factorized as $P_0(O) = P_0(Y|\mathbf{A}, W)P_0(\mathbf{A}|W)P_0(W)$. We use $p(\cdot)$ to denote the corresponding densities, e.g. $p(Y|A,W)$. The conditional joint density of exposures given covariates $p(\mathbf A | W)$ we will refer to as \(g_0(\mathbf{A}|W)\) to distinguish it. Similarly we use \(\bar{Q}_0(\mathbf{A}, W) \equiv E_0(Y|\mathbf{A}, W)\). We denote a ``shifted'' exposure density using the notation $ g_\delta(\mathbf{A}|W) = g_0(\mathbf{A}-\boldsymbol{\delta}|W)$ and we denote the full ``shifted'' density over all variables as $p_\delta(O) = p(Y|\mathbf A, W) g_\delta (\mathbf A|W) p(W)$. In this section, while we theoretically describe the shifts \(\boldsymbol{\delta}\) as fixed quantities (in particular, which variables are shifted at all and in what amounts), practical applications require them to be identified in a data-adaptive manner. We first describe the parameter that we will estimate treating exposure shifts as fixed, and then we describe how to identify what exposures to shift and by how much. 

For example, consider $A = (A_1, A_2)$ (e.g., $A_1$ = dioxin, $A_2$ = furan, both POPs), here we are interested in the probability of the outcome given joint exposure. We are then interested in reducing this exposure profile by $\boldsymbol{\delta} = (\delta_1, \delta_2)$ which can be static (say some practical unit or standard deviation) or this can be a function \(\boldsymbol{\delta}(W)\), that is, a change amount based on the history of the covariates. We use $P$ to denote a generic probability distribution within a nonparametric statistical model $\mathcal M$. The symbol $P_n$ represents the empirical distribution of the observed data, attributing a probability of 1/n to each observation for the sample size $n$. The mapping of target parameters is defined as $P \rightarrow \Psi(P)$, with $\Psi(P_0)$ as the true target parameter under $P_0$, and $\hat{\Psi}(P_n)$ as an estimator mapping the empirical distribution $P_n$ to its estimated value. The true value of the target parameter is $\psi_0 = \Psi(P_0)$, and its estimate based on $P_n$ is represented as $\psi_n$. In causal inference, we often seek to contrast observed data with potential outcomes under different intervention scenarios (the full causal model). We begin by first clarifying the distinction between statistical and causal parameters for stochastic shift interventions for potentially multiple exposures, then we move on to build our parameter from this identification. We first define our causal parameter using counterfactuals \cite{Rubin1974-od, Rubin2005-by}. For each unit in our population, we have an infinite family of potential outcomes \( Y(\cdot) \) corresponding to every possible value of \( \mathbf A \). This family, \( Y(\cdot) \), is considered a stochastic process. In practice, for each unit, we observe only one of these potential outcomes, specifically \( Y(\mathbf A) \), the outcome corresponding to the observed treatment. However, we can define a true \textit{causal} distribution $P^*$ as a distribution over the variables $Y(\cdot), \mathbf A, W$. The condition $Y = Y(\mathbf A)$ defines a mapping $\mathcal O$ from causal distributions $P^*$ to observable distributions $P$. Our causal goal is to define potential outcomes via intervention on \(\mathbf A\) to a new value, \(\mathbf A - \boldsymbol{\delta}\). In this context, our causal parameter is defined with respect to the causal distribution $P^*$: 
$$\Psi^*(P^*) = \mathbb{E}_{P^*}[Y(\mathbf{A}-\boldsymbol{\delta})]$$
However, because $P^*$ includes unobserved potential outcomes, we cannot estimate \(\Psi^*(P^*)\) directly. Instead, we rely on identification strategies that allow us to express the causal parameter \(\Psi^*(P^*)\) in terms of the observed data distribution, thereby allowing us to make a practical estimation from the data we have. The assumptions needed for this are 
conditional ignorability: \( Y(a) \perp A | W \) for all \( a \) and positivity: $ 0< g_{\delta}(A|W) / g(A|W) <M  $ with probability 1. With these assumptions, we can identify our causal parameter via the following statistical parameter:

$$\Psi(P) = E_{P_\delta}[Y] = \int_{\mathbf{A}} \int_{W} \bar Q (\mathbf{a}, w) g_\delta(\mathbf{a}|w) \, p(w) \, d\mathbf{a} \, dw$$

This captures the distribution of outcomes \( Y \) if we were to simply change the observed exposure \( \mathbf A \) by some amount $\boldsymbol{\delta}$, while keeping all other factors unchanged. Throughout, let $E_\delta[\cdot]$ abbreviate $E_{P_\delta}[\cdot]$. Note that despite being defined in terms of $P_\delta$, $\Psi(P)$ is indeed a function of $P$ because $P_\delta$ is defined in terms of $P$. Thus, upon invoking these assumptions,$\Psi^*(P^*) = \Psi(\mathcal O(P^*)) = \Psi(P)$. From now on, we focus on the target parameter $\Psi$ (defined on the observable distribution) and our observable estimand $\psi_0$ (the value $\Psi$ takes at our true, unknown distribution $P_0$). In practice, identifying causal effects requires thorough consideration of these assumptions. The conditional ignorability assumption, in particular, asserts that there are no unmeasured confounders influencing both the exposure and the outcome, which is a substantial claim, especially in observational data where mixed exposures are usually observed. The positivity assumption ensures that there is enough variability in our data to make meaningful causal inferences. When these assumptions are reasonable and satisfied, the statistical parameters derived from the observed data can be insightful proxies for the true causal parameters. Alternatively, the resulting quantities can be interpreted as pure statistical parameters of interest, a particular type of association of $A$ with $Y$ controlling for $W$, \cite{Van_der_Laan2006-ro}. In this way, we can interpret ``causal parameters '' as a variable importance measure. 

\subsection{Interaction Target Parameter}

Our target parameter above defines the causal parameter for an arbitrary joint shift of multiple exposures under the stochastic intervention framework. Realistically, given the usual size and complexity of public health data, no more than two-way interactions can normally be reliably determined. As such, we focus on describing the interaction parameter for two exposures that builds off the joint stochastic shift parameter previously discussed; it is straightforward to generalize the above to more than two exposures. The symbol $\boldsymbol{\delta}$ in bold represents the shift of both exposures by two shifted amounts, $E_{\delta_i} [Y]$ represents the change in $A_i$ by $\delta_i$ leaving $A_j$ at the observed levels, and $E_{\delta_j} [Y]$ represents a shift in $A_j$ by $\delta_j$ leaving $A_i$ at the observed levels. As such, we define our bivariate interaction parameter: 
$$E_{\boldsymbol{\delta}} [Y] - \left( E_{\delta_i}[Y] + E_{\delta_j}[Y] \right) + E[Y]$$. 
Which is the expected outcome under the joint shift of $\mathbf{A}$ compared to the expected additive outcome under each individual shift in $\mathbf{A}$. The interpretation of this parameter is similar to that of an interaction term in a linear model. 

\section{Estimation and Interpretation}

\subsection{Efficient Influence Function}

In expanding the scope of Diaz and van der Laan's work on TMLE for these shift intervention parameters \cite{diaz_2012} to consider exposures as vectors rather than scalar entities, we simply modify the notation to possible sets of exposures accordingly. The EIF for $E_{\boldsymbol{\delta}}[Y]$ is given by $D_{\boldsymbol{\delta}}(P_0)(o) = H_{\boldsymbol{\delta}}(\mathbf{a}, w)({y - \overline{Q}(\mathbf{a}, w)}) +
\overline{Q}(\mathbf{a}- \boldsymbol{\delta}, w) - E_{\boldsymbol{\delta}}[Y]$. Where \( H_{\boldsymbol{\delta}} \) is defined as $H_{\boldsymbol{\delta}}(\mathbf{a},w) = g_0(\mathbf{a} - \boldsymbol{\delta} \mid w)/
g_0(\mathbf{a} \mid w)$. Note that vectors \( \mathbf{a} \) and \( \boldsymbol{\delta} \) emphasize the element-wise operations and comparisons. Extension to a vectorial representation allows us to use the original EIF for multiple exposures. The auxiliary covariate \( H_{\boldsymbol{\delta}}(\mathbf{a},w) \) serves as a ratio of conditional (possibly joint) densities \( p(\mathbf{a}|W) \). Specifically, \( g_0(\mathbf{a} \mid w) \) denotes the conditional density under observed values, while \( g_0(\mathbf{a} - \boldsymbol{\delta} \mid w) \) is the density after a decrease of \( \boldsymbol{\delta} \) to \( \mathbf{a} \). This ratio captures the density's change upon a \( \boldsymbol{\delta} \) shift. In our approach, we use TMLE to remove the plug-in bias of our \(\bar{Q}\) estimators. Given the interaction parameter defined as $\Psi(P) = E_{\boldsymbol{\delta}} [Y] - \left( E_{\delta_i}[Y] + E_{\delta_j}[Y] \right) + E[Y]$, the EIF for this interaction can be succinctly represented in terms of the EIFs for the individual components 
$D(P) = D_{\boldsymbol{\delta}}(P) - \left( D_{\delta_i}(P) + D_{\delta_j}(P) \right)$, where $D_{\boldsymbol{\delta}}(P)$ is the EIF of $E_{\boldsymbol{\delta}} [Y]$ and $D_{\delta_i}(P)$ and $D_{\delta_j}(P)$ are the EIFs of $E_{\delta_i}[Y]$ and $E_{\delta_j}[Y]$, respectively.

\subsection{TMLE}

We chose to estimate our parameter using TMLE (rather than estimating equations, inverse propensity of treatment weights (IPTW) and augmented IPTW ) chiefly due to its favorable finite sample properties due to being a plug-in estimator \cite{best_tmle}. Briefly, to construct the TMLE estimator, we: 1. Use data-adaptive regression for the initial estimation of \( g_0(\mathbf{a}, W) \) and \( \overline{Q}_0(\mathbf{a}, W) \), 2. Evaluate \( H_\delta(\mathbf{a}_i,w_i) \) for each observation, 3. Using the derived estimates, we implement a logistic regression to refine  predictions under the shift $\text{logit}\overline{Q}_{\epsilon, n}(\mathbf{a} - \delta, w) = \text{logit}\overline{Q}_n(\mathbf{a} - \delta, w) + \epsilon H_{\delta,n}(\mathbf{a}, w)$ then 4. From this model, once $\epsilon $ is determined, we update the predicted counterfactuals, using these updated counterfactuals, we then compute the plugin. This is a standard TMLE procedure for stochastic interventions; we simply shift multiple exposures simultaneously in the initial estimation and update. We estimate a standard error by taking the empirical variance of the estimated EIF and use this construct Wald-type confidence intervals and p-values. We use TMLE to ascertain unbiased estimates for each part of our interaction target parameter. That is, in $E_{\boldsymbol{\delta}} [Y]$ we update our counterfactual outcome under a joint shift by the joint density ratio as $H_\delta(\boldsymbol{a}, w)$. Similarly, $E_{\delta_1}[Y]$ and $E_{\delta_2}[Y]$ are simply updated based on $H_\delta(a,w)$ for univariate shifts. In general, we use TMLE to target each of the three parts of our parameter. Given that our interaction is a continuous function (linear combination) of these asymptotically efficient estimates, we use the delta method for variance derivation.

\section{Exposure Set Identification}

The above presentation of the estimation of the marginal, joint and interaction shift parameters assumes that the subsets of the mixture that are important with respect to the outcome, as well as the pairs that have evidence of statistical interaction, are known a priori. However, in finite samples with a large number of mixture variables, $\mathbf{A}$, including all components of the mixture as potentially impactful exposures and all possible interactions will result in unacceptable sample variability (and potential nonnormality). Therefore, we propose using a data-adaptive target parameter approach \cite{Hubbard2016} to use sample splitting to ``discover'' the relevant components and interactions among the mixture variables and estimate the resulting data-adaptively defined parameters.

We propose a K-fold cross-validation in which the sample data are split into K-folds of roughly equal size. For example, consider $K$ = 10, and define $k = 1$. The $K-1 /K$ portion of the sample is used to define the parameter of interest (here 90\%) and the $1/K$ (10\%) is used to estimate the parameters. Let $T_{n,k} \subset \{1, \ldots, n\}$ denote the observations indices in the k-th training sample and $V_{n,k}$ denote the observations indices in the k-th validation sample for $k = 1, \ldots, K$. The following paragraphs describe how through cross-validation the $T_{n,k}$ (training) portion of the data is used to identify important variable sets in the mixture and $V_{n,k}$ (validation) is used to estimate our stochastic shift intervention parameters. 

\subsection{Identifying Interactions in the Mixture Using Training Data}

The discovery of important subsets of variables and interactions in the mixture variables is done using an ensemble method: the discrete Super Learner (SL) \cite{SL_2008} and g-computation \cite{Robins1987, Daniel2013}. The SL algorithm employs a cross-validation mechanism, optimizing over a library of candidate algorithms to select the estimator with the best empirical fit. We use a discrete SL to select the best fitting estimator from a library of different estimators for the estimation of $\mathbb{E} [Y | \boldsymbol{A}, W]$. Using this SL, we then predict $\mathbb{E} [ \mathbb{E}[Y | A_i - \delta_i, \textbf{A}_{-i}, W]]$, which is the individual exposure effects for exposure $A_i$, we do this for each exposure in the mixture. For all 2-way combinations, we then calculate: $\mathbb{E}[ \mathbb{E} [Y | A_i - \delta_i, A_j - \delta_j, \mathbf{A}_{- i,j}, W]]$, the expected outcome of a joint shift for all 2-way combinations. By comparing the estimated outcome under joint shift compared to the expected outcome sum of individual shifts for the same exposures and $\delta$'s we get a g-computation estimate for our interaction parameter, $\Psi_{\text{IE}}$, using the training data. 

\subsection{Ranking Synergistic and Antagonistic Interactions}

For our purposes, we want to rank interaction effects based on their magnitude to identify the most significant synergistic and antagonistic effects, these define our oracle parameter sets of interactions:
  \begin{align*}
    \mathcal{S}_s &= \{ (i, j) | \text{ranked by } \Psi_{\text{IE}, ij} \text{ in descending order}, \forall i < j \}[:s] \\
    \mathcal{A}_s &= \{ (i, j) | \text{ranked by } \Psi_{\text{IE}, ij} \text{ in ascending order}, \forall i < j \}[:s]
    \end{align*}

Where \(\mathcal{S}_s\): are the top \(s\) synergistic effects and \(\mathcal{A}_s \) are the top \(s\) antagonistic effects. This is because, if the joint impact is much greater than the additive impact for a change in exposures $i,j$, then this indicates a synergistic or superadditive relationship (the change in both is much more than additive). Inversely, if the interaction parameter is negative, this indicates that a change in both is less than the additive effect, which implies antagonism between the exposures. 

In this framework, we have a family of parameters indexed by \( (i,j) \), and we are particularly interested in the parameter \( \Psi_{i_0,j_0}(P) \) where \( (i_0,j_0) \) maximizes the effect size over all possible pairs \( (i,j) \). This maximizing parameter embodies the oracle parameter and helps us understand which interactions demonstrate the greatest synergy (most positive effect) or antagonism (most negative effect).

Defining these oracle parameters as the top synergistic/antagonistic interactions is imperative as they describe a consistent parameter that can be estimated in every fold of the CV procedure. That is, if we are simply interested in the max and min interaction effect (top synergistic and antagonistic interactions) this is estimable in every partition of the data, we can then pool our results for this parameter, and if the exposure sets used in the top ranks are consistent, we gain power for the interaction parameter and thus have tighter confidence intervals. Thus, our oracle target parameters are the top subsets of synergistic and antagonistic interactions based on our interaction parameter definition. The number of subsets, such as top 1 or top 3 ranked subsets are predefined by the user in our package. 

\subsection{Estimation of Shift Interventions on Discovered Variable Sets}

Once these exposure sets for the top synergistic and antagonistic interactions are selected in the training data, we treat these exposure sets as fixed, we then estimate the resulting data-adaptively defined parameters using TMLE for marginal and joint stochastic shift interventions used to construct the interaction parameter.  Using the training data, the nuisance functions \(g_n\) and \(Q_n\) are trained. The TMLE update step is performed using these training estimates to estimate $\epsilon$. With the validation data, we obtain estimates for \(g_n\) and \(Q_n\) and apply $\epsilon$ to debias our initial plug-in for our estimated stochastic shift target parameter using the validation data. In this case, the validation data are data that correspond to $k = 1$. This procedure describes CV-TMLE \cite{hubbard_book_da, Hubbard2016} which is the infrastructure we use to estimate our target parameter. Readers interested in the details of this procedure are directed to these resources.

With the estimates for our parameter complete for the validation for $k = 1$, we would then move on to $k = 2$ and repeat the sections 4.1-4.3. Upon completion of the cross-estimation procedure, a pooled TMLE update provides a summary measure of the target parameter across k-folds. Specifically, we stack the predictions for each nuisance parameter from the estimation samples and run a pooled TMLE update on the cumulative initial estimates within each interaction rank (for example, the top synergistic interaction across all the folds). The resulting average is then used for parameter estimation. The pooled estimate across the folds can be succinctly represented as:

$$\theta = \frac{1}{K} \sum_k \Psi_{F_{T_{n,k}}}\left(V_{n,k}\right)(P_0)$$
Where $\Psi_{F_{T_{n,k}}}$ identifies optimal exposure sets and subsequently formulates a plug-in estimator using training data $ T_{n,k} $. $ V_{n,k} $ refers to the estimation sample of the data for deriving our estimates.

\subsection{Pooled Estimates under Data-Adaptive Delta}

In stochastic interventions, the selection of shift magnitude \(\boldsymbol{\delta}\) is critical, especially in the presence of joint stochastic shifts. These shifts can lead to positivity violations, which may bias and inflate the variance of the estimated exposure effects. To mitigate these risks, we employ a data-adaptive strategy where \(\delta\) is dynamically adjusted within each training fold to maintain the ratio \(H_\delta(a,w)_i = g_{n_{-k}}(a_{n_{-k}} - \delta \mid w)/g_{n_{-k}}(a_{n_{-k}} \mid w)\) below a pre-specified threshold \(\lambda\) (default is 50 in our implementation). This ensures that all shifts are supported by the data, minimizing the risk of extreme or unsupported extrapolations.

When \(\delta\) varies across folds due to differences in the data distributions or density estimates, we face the challenge of pooling results to form a coherent overall estimate. Instead of selecting a single \(\delta\) such as the minimum or maximum, we opt to average the deltas. This approach is taken to reflect a shift scenario that is, on average, supported across all subsets of the data, thereby representing a more robust and generalizable intervention scenario. We then stack the nuisance parameters corresponding to these varied \(\delta\) values and perform a TMLE update across the entire pooled dataset. We then plug these debiased counterfactuals into our influence function using the full data across various shifts. The resulting estimator reflects the average causal effect under these averaged shift scenarios.

This method acknowledges that, while individual fold-specific deltas might capture local data characteristics, the averaged delta provides a balanced view that is likely to be more stable and representative of the population as a whole. Nevertheless, this approach introduces a layer of complexity in interpretation, as the estimated effects correspond to an 'average' intervention rather than a specific fixed intervention.

\section{Density Estimation of the Exposures}

Estimating the density of each exposure is crucial for evaluating the effects of interventions both on average across the population and within subpopulations. We provide two approaches to density estimation: direct density estimation and reparameterization based on classification, depending on user input.

\subsection{Direct Density Estimation}

Super Learner Algorithm: The Super Learner algorithm operates by training a variety of base learners (individual algorithms) on the data and then combining their predictions using a meta-learner. The meta-learner, in this case, is a non-negative least squares (NNLS) algorithm, which assigns weights to each base learner based on their performance. This approach leverages the strengths of different algorithms to produce a more accurate and robust predictive model.

Homoscedastic Errors (HOSE) Learner: The HOSE learner assumes that the variance of the errors is constant (homoscedasticity) across the range of predicted values. It employs a semi-parametric approach to density estimation, where the mean of the distribution is estimated using a Super Learner composed of multiple base learners.

Heteroscedastic Errors (HESE) Learner: The HESE learner, in contrast, allows for varying error variances (heteroscedasticity) across different levels of predicted values. It employs a two-step semi-parametric approach. The mean of the distribution is estimated using a similar Super Learner as in HOSE. An additional Super Learner is used to model the variance of the errors, allowing the error variance to change depending on the value of the predictors.

Although direct estimation is straightforward in principle, it can be challenging in practice due to the scarcity of effective data-adaptive estimators for high-dimensional conditional densities in the machine learning and statistical literature. As such, we also provide an alternative approach.

\subsection{Reparameterization of Density Estimation via Classification}

As an alternative to direct density estimation, we propose a novel approach that redefines the problem of estimating the density ratio \( H(\mathbf{A}, W) \) as a classification problem in an augmented dataset. This method leverages the fact that classification algorithms are well-developed and widely available in the machine learning literature compared to density estimators. This approach is used in other methodologies such as in longitudinal modified treatment policies \cite{Diaz2023}. 

To apply this approach, we construct an augmented dataset consisting of duplicated observations. For each original observation, we create a duplicate with the observed exposures \(\mathbf{A}\) and another with the exposures under the intervention \(\mathbf{A}^{\delta}\). We introduce an indicator variable \(\xi\), which is set to 1 if the duplicated observation corresponds to the exposure under intervention, and 0 otherwise. Here, exposure can be the intervention on multiple exposures, such as the joint shift we need in our interaction parameter. The augmented dataset can be represented as:

\[ \{(W_{\xi,i,t}, \mathbf{A}_{\xi,i,t}, \xi_{\xi,i}): \xi = 0, 1; i = 1, \ldots, n\}, \]

where \(\xi_{\xi,i} = \xi \) indexes the duplicates, \( W_{\xi,i,t} = W_{i,t} \) is the history variable, and \( \mathbf{A}_{\xi,i,t} = \xi \cdot \mathbf{A}^{\delta}_{i,t} + (1 - \xi) \cdot \mathbf{A}_{i,t} \) is the natural exposure level if \(\xi = 0\), and the intervened exposure level if \(\xi = 1\).

We denote the probability distribution of \( (W_t, \mathbf{A}_t, \xi) \) in the augmented dataset by \( P_\xi \), with corresponding density \( p_\xi \). We define the parameter \( u_\xi \) of \( P_\xi \) as:

\[ u_\xi (\mathbf{a}_t, w_t) = P_\xi(\xi = 1 \mid \mathbf{A}_t = \mathbf{a}_t, W_t = w_t). \]

The density ratio \( H(\mathbf{A},W) \) can be expressed in terms of \( u_\xi \) as follows:

\[ H_t(\mathbf{a}_t, w_t) = \frac{p_\xi(\mathbf{a}_t, w_t \mid \xi = 1)}{p_\xi(\mathbf{a}_t, w_t \mid \xi = 0)} = \frac{P_\xi(\xi = 1 \mid \mathbf{A}_t = \mathbf{a}_t, W_t = w_t)}{P_\xi(\xi = 0 \mid \mathbf{A}_t = \mathbf{a}_t, W_t = w_t)} = \frac{u_\xi (\mathbf{a}_t, w_t)}{1 - u_\xi (\mathbf{a}_t, w_t)}, \]

where the first equality follows from the definition of \( H_t \) and the definition of conditional density, the second from Bayes' rule and the observation that \( P_\xi(\xi = 1) = P_\xi(\xi = 0) = \frac{1}{2} \), and the third from the definition of \( u_\xi \).

By estimating \( u_\xi \) in the augmented dataset using any classification method available in the machine learning and statistical learning literature (e.g., Super Learner), we can directly estimate the density ratio. This classification-based approach ensures that the estimator retains desirable properties such as asymptotic normality, provided that appropriate cross-validation and cross-fitting techniques are employed. Furthermore, this reclassification approach is substantially faster computationally, which is necessary, especially in our proposed methodology, where the conditional density is needed for each exposure and higher folds in cross-validation provide more consistent results. Overall, this approach has been used in previous literature to avoid density estimation. We simply create a duplicate data set, shift exposures by $\delta$ and create an intervention outcome = 1 for this data set. Bind rows with our unperturbed data with the outcome intervention = 0. Then we train a classifier and use Bayes rule to estimate the density ratio comparing shift to no shift. 

In our implementation, users can choose between direct density estimation and the reparameterization approach via a user-defined parameter. Moving forward, all applications and simulations use the classification approach.

\section{Simulations}

In this section, we present simulations to demonstrate the effectiveness of our proposed method in identifying the correct exposures and interactions used to generate the outcome. The data-generating process (DGP) is designed to reflect the complexity of mixed exposures, where mixture variables are correlated but only some have a significant impact on the outcome.

\subsection{Data-Generating Process}

We consider a random vector of covariates, $\mathbf{W} = (W_1, W_2, W_3, W_4, W_5, W_6, W_7)$, generated from a multivariate normal distribution with mean vector $\boldsymbol{\mu} = (0, 0, 0, 0, 0, 0, 0)$ and a covariance matrix $\boldsymbol{\Sigma}_W$. The covariance matrix $\boldsymbol{\Sigma}_W$ is designed to reflect real-world dependencies among covariates such as age, sex, BMI, systolic blood pressure (SBP), diastolic blood pressure (DBP), cholesterol, and fasting blood sugar (FBS).

Based on these covariates, we generate the exposures $\mathbf{A} = (M_1, M_2, M_3, M_4, M_5, M_6)$. Specifically, $M_1$ is a lognormal variable influenced by age ($W_1$) and sex ($W_2$): $M_1 = \exp(0.2 W_1 + 0.1 W_2 + \epsilon)$, where $\epsilon \sim \mathcal{N}(0, 0.5)$. $M_2$ is a gamma variable influenced by BMI ($W_3$) and another variable ($W_4$): $M_2 \sim \text{Gamma}(2, \exp(0.3 W_3 + 0.2 W_4))$. $M_3$ is a beta variable influenced by DBP ($W_5$) and random noise: $M_3 \sim \text{Beta}(1 + 0.1 W_5, 2 + 0.2 \epsilon)$, where $\epsilon \sim \mathcal{N}(0, 1)$. $M_4$ is a mixture of normal and uniform distributions (bimodal): $M_4 = \mathbb{I}(\text{Uniform}(0,1) > 0.7) \cdot \mathcal{N}(0.5 W_4, 0.3) + \mathbb{I}(\text{Uniform}(0,1) \leq 0.7) \cdot \text{Uniform}(0, 1)$. $M_5$ is a truncated normal variable influenced by cholesterol ($W_6$) and FBS ($W_7$): $M_5 \sim \text{TruncNormal}(\text{min}(W_6, \text{quantile}(W_6, 0.05)), \text{max}(W_6, \text{quantile}(W_6, 0.95)), 0.1 W_6 + 0.2 W_7, 0.4)$. $M_6$ is a mixture of normal distributions with right skew (lognormal-like): $M_6 = \mathbb{I}(\text{Uniform}(0,1) > 0.8) \cdot \mathcal{N}(0.2 W_5 + 0.1 W_6, 0.2) + \mathbb{I}(\text{Uniform}(0,1) \leq 0.8) \cdot (\mathcal{N}(0.2 W_5 + 0.1 W_6, 0.1) + \mathcal{N}(0, 1))$.

The outcome variable $Y$ is then generated by incorporating both synergistic and antagonistic interactions among exposures: $Y = 3 + 1.5 M_1 + 1.2 M_2 - 0.8 M_3 + 2 (M_1 \times M_2) + M_4 + M_5 + \epsilon$, where $\epsilon \sim \mathcal{N}(0, 1)$ is a random noise term. Here, $M_1 \times M_2$ represents a synergistic interaction that becomes stronger as the concentrations increase. Our goal is to accurately detect and estimate this synergistic interaction parameter as well as each part of it (the marginal and joint shifts used to construct the interaction parameter).

\subsection{Analysis of Shifts and Interactions}

We generate a large dataset $P_0$ of 100,000 units. For each pair of exposures, we apply shifts of 1 unit to each exposure and calculate the effects of single and pairwise shifts on the outcome. We compute the true difference between the combined effect of shifting both exposures and the sum of their individual effects to quantify the interactions. Specifically, we are interested in the following true effects built into the DGP: the top synergistic effect, which is 2.012, representing the average change in outcome beyond the additive effect (joint shift = 11.35 in $M_1$ and $M_2$ compared to the sum of individual shifts in these variables, $M_1 = 5.81$, $M_2 = 3.53$); and the top positive effect for a shift in $M_1$, which is 5.81.

These represent the true values we want to compare our estimates to. We run our InterXshift algorithm, which attempts to find the top synergistic, joint and marginal results. The goal is to assess the performance of our method in accurately identifying and estimating the impacts of synergistic interactions in a realistic mixed exposure scenario.

\subsection{Evaluating Performance}

We evaluated the asymptotic convergence to the true counterfactual differences for these exposures, in each simulation, which inherently means convergence to identifying the true exposure relationships (if bias converges to 0 for the top synergy relationship, then this means we are both finding the true synergistic relationship and correctly estimating our interaction parameter). To do so, we followed the following steps 1. From our $P_0$ population, we generated a random sample of size $n$ from the DGP, 2. At each iteration, we used the training sample to define the exposure(s) and create the necessary estimators for the target parameter dependent on the exposure sets. We then used the validation sample to obtain the updated causal parameter estimate using TMLE. We repeated this process for all the folds, 3. At each iteration, we output the stochastic shift estimates given the pooled TMLE for each effect, the top synergistic relationship, top antagonistic relationship, top positive marginal relationship, and top inverse marginal relationship.

To evaluate the performance of our approach, we calculated several metrics for each iteration, including bias, variance, MSE, and confidence interval (CI) coverage. To visually inspect if the rate of convergence was at least as fast as $\sqrt{n}$, we show absolute bias and MSE times $\sqrt{n}$.  We then calculated the variance for each estimate and used it to calculate the mean squared error (MSE) as $\text{MSE} = \text{bias}^2 + \text{variance}$. For each estimate, we calculated the CI coverage as the proportion of iterations the true parameter was covered by estimated confidence intervals.  We calculated these performance metrics at each iteration, performing 100 iterations for each sample size $n =$ (300, 600, 1000, 2000, 3000). We used \texttt{InterXshift} with 5-fold cross-validation (to speed up calculations in the simulations) and default learner stacks (see below) for each nuisance parameter and data-adaptive parameter. 

To use \texttt{InterXshift}, we need estimators for $\bar{Q} = E(Y|A,W)$ and $\xi$ (density ratio learner). \texttt{InterXshift} provides default algorithms to be used in a Super Learner \cite{SL_2008} that are both fast and flexible. To estimate $\bar{Q}$, we include estimators for the Super Learner from glm, elastic net \cite{elasticnet}, random forest \cite{ranger}, and xgboost \cite{xgboost}. The same learners are used for our $\xi$ learner.

\subsection{Results}

\begin{figure*}
  \centerline{\includegraphics[width=7.0in]{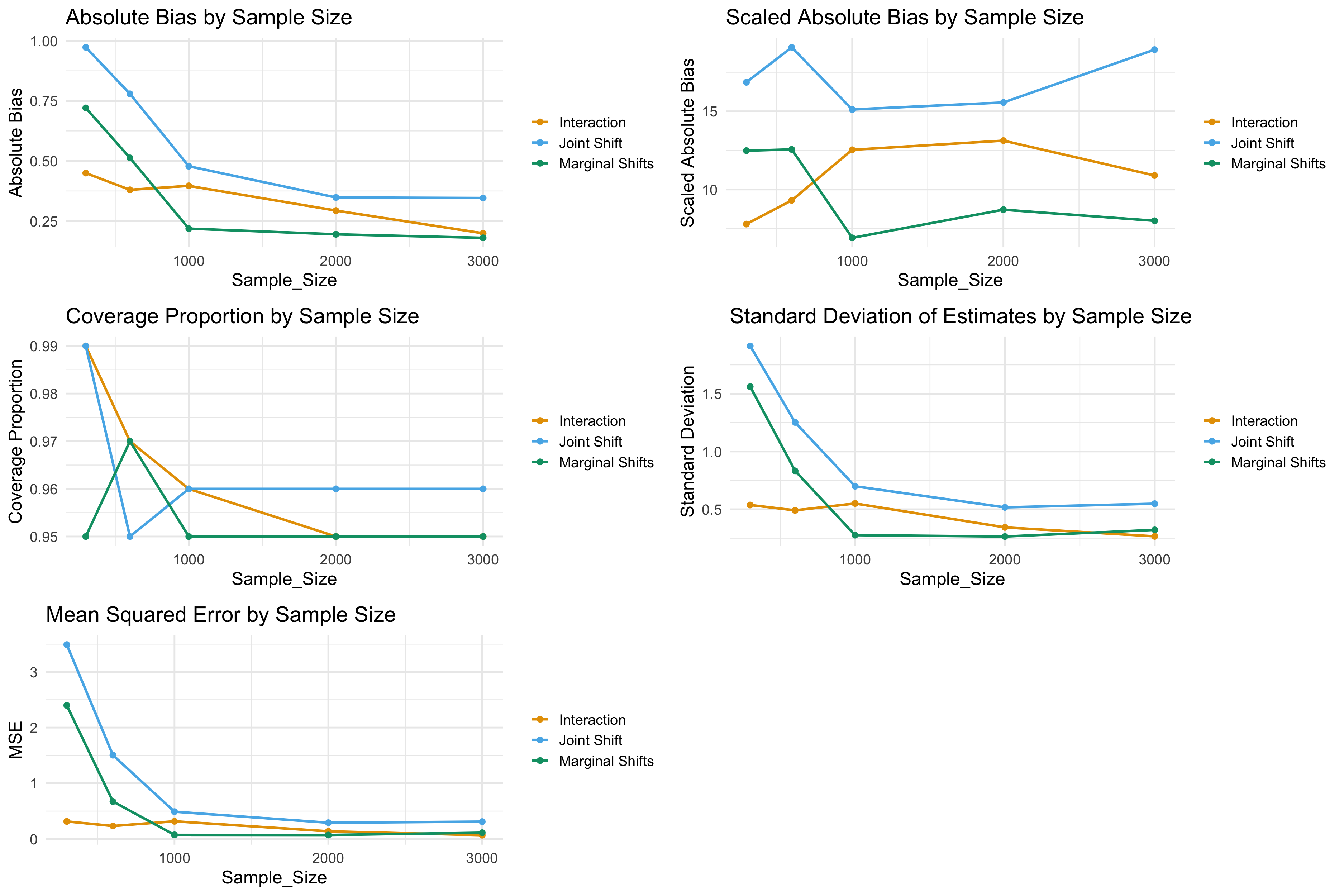}}
  \caption{Absolute Bias, Absolute Bias times $\sqrt{n}$, Proportion CI Coverage, Standard Deviation and MSE for Each Parameter Across Sample Sizes: These metrics represent results from CV-TMLE, that is bias is w.r.t. the oracle target parameter where we use CV to identify the estimated top ranked synergistic interaction then estimate the joint and marginal shifts for our interaction parameter.}
\label{fig:convergence}
\end{figure*}

\textbf{InterXshift consistently discovers and estimates the correct interaction}. The top synergistic interaction effect was identified in 100\% of the folds in all sample sizes.  \textbf{Figure \ref{fig:convergence}} shows the absolute bias, bias times $\sqrt{n}$, CI coverage, mean squared error, and standard deviation as the sample size increases to 3000. For bias, MSE and standard deviation, there is a converge to near zero at sample size 3000. For coverage, the average coverage for each target parameter was: individual shift: 95.5\%; joint shift: 96\%; interaction: 96\%. Coverage was slightly high at lower sample sizes but this is likely due to variability in the simulation. The absolute bias times $\sqrt{n}$ should look flat if convergence is at $\sqrt{n}$ rate which we generally see with some variability in the simulation. 

\section{Applications}

\subsection{Analysis of NIEHS Synthetic Mixtures Data}

The NIEHS synthetic mixtures data is a commonly used data set to evaluate the performance of statistical methods for mixtures. These synthetic data can be considered the results of a prospective cohort study, where the outcome cannot cause exposures, and correlations between exposure variables can be thought of as caused by common sources or modes of exposure. The variable $W$ can be assumed to be a potential confounder and not a collider. The data set has 7 exposures ($A_1 - A_7$) with a complex dependency structure based on endocrine disruption. Two groups of exposure ($A_1, A_2, A_3$ and $A_5, A_6$) lead to high correlations within each group. $A_1, A_2, A_7$ positively contribute to the outcome, $A_4, A_5$ have negative contributions, while $A_3$ and $A_6$ have no impact on the outcome. Rejecting $A_3$ and $A_6$ is difficult due to their correlations with the members of the cluster group. This correlation and effects structure is biologically plausible, as different congeners of a group of compounds may be highly correlated but have different biological effects. Exposures have various agonistic and antagonistic interactions, a breakdown of the variables sets and their relationships are: $A_1, A_2$ both increase Y by concetration addition, $A_1 - A_4, A_2 - A_4, A_1 - A_4, A_2 - A_4$ competetive antagonism, $A_1-A_7, A_2-A_7$ supra-additive or synergistic, $A_4, A_5$ both addivively decrease Y, $A_4-A_7, A_5-A_7$ unusual antagonism. Synthetic data and the key for dataset 1 are available on \href{https://github.com/niehs-prime/2015-NIEHS-MIxtures-Workshop}{GitHub}. This resource shows the interactions and marginal dose relationships built into the data. Given these toxicological interactions, we expect these sets of variables to be determined in \texttt{InterXshift}. For example, we might expect a positive counterfactual result for A1, A2, A7 and negative results for A4 and A5. \textbf{Table \ref{tab:NIEHS_intxns}} shows the interactions built into this data.

\begin{table}[h]
\centering
\begin{tabularx}{\textwidth}{l|X} 
\toprule
Variables & Interaction Type                                                                          \\ 
\midrule
A1 and A2 & Toxic equivalency factor, a special case of concentration addition (both increase Y) \\
A1 and A4 & Competitive antagonism (similarly for A2 and A4)                                          \\
A1 and A5 & Competitive antagonism (similarly for A2 and A4)                                          \\
A1 and A7 & Supra-additive (“synergy”) (similarly for A2 and A7)                                      \\
A4 and A5 & Toxic equivalency factor, a type of concentration addition (both decrease y)              \\
A4 and A7 & Antagonism (unusual kind) (similarly for A5 and A7) \\ 
\bottomrule                                     
\end{tabularx}
\caption{NIEHS Synthetic Data Interactions}
\label{tab:NIEHS_intxns}
\end{table}

Likewise, in the case for antagonistic relationships such as A1-A5 (third row in \ref{tab:NIEHS_intxns}), we would expect a joint shift to come closer to the null, since A5 antagonizes the positive effects of A1. For A1 and A2, we would expect the joint shift to be close to the sum of individual shifts (not much interaction), but for A1 and A7 we expect to estimate more than an additive effect (some interaction). The NIEHS data set has 500 observations and 9 variables. $W$ is a binary confounder. Based on this table, we can gauge \texttt{InterXshift}'s performance by determining if the correct synergistic and antagonistic relationship are found and if the correct variables are rejected. 

We apply \texttt{InterXshift} to these NIEHS synthetic data using a three-fold CV (smaller CV to save space here in k-fold result tables). We use the default libraries of learners for the estimation of $\xi$ and $Q$.  We parallelize over the cross-validation to test computational run-time on a newer personal machine an analyst might be using. We use a delta of 1 for all the exposures in the mixture. We present the pooled TMLE results and the k-fold specific results.

\subsection{NIEHS Synthetic Data Results}
\texttt{InterXshift} accurately rejects exposures $A_3, A_6$, finding the correct top positive marginal effects of $A_1$ in all the folds and the top inverse association ($A_5$) in all the folds. Likewise, we found the correct synergistic results built into the DGP for $A_1-A_7$ and $A_2-A_7$. In all the folds, $A_5-A_7$ was found to have the strongest antagonistic relationship, which is also true based on the DGP key provided from the NIEHS. 

\textbf{Table \ref{tab:rank1_pos}} presents the rank 1 positive marginal results for the NIEHS data. The first three rows display fold-specific results, with $\Psi$ representing the expected change in the outcome under the exposure shift compared to the average observed outcome. Here, $A_1$ is identified as the top positively associated mixture variable across all folds. The pooled TMLE column provides the aggregate results, showing an average $\Psi$ of 16.31 with a standard error of 0.37 and a 95\% confidence interval of 15.59 to 17.04. These results underscore the consistent positive association of $A_1$ with the outcome across different data folds.

\begin{table}[ht]
\hspace{-1cm}\begin{tabular}{rlrrrrrrlrr}
  \hline
 & Condition & $\Psi$ & Variance & SE & Lower CI & Upper CI & P-value & Fold & N & Delta \\ 
  \hline
1 & A1 & 20.66 & 0.45 & 0.67 & 19.35 & 21.98 & 0.00 & 1 & 167 & 1.00 \\ 
2 & A1 & -2.66 & 1.28 & 1.13 & -4.88 & -0.44 & 0.02 & 2 & 167 & 1.00 \\ 
3 & A1 & 19.13 & 0.53 & 0.73 & 17.70 & 20.56 & 0.00 & 3 & 166 & 1.00 \\ 
4 & Rank 1 Pos & 16.31 & 0.14 & 0.37 & 15.59 & 17.04 & 0.00 & Pooled TMLE & 500 & 1.00 \\ 
   \hline
\end{tabular}
\caption{Rank 1 Positive Marginal Results for NIEHS Data}
\label{tab:rank1_pos}
\end{table}

\begin{table}[ht]
\hspace{-1cm}\begin{tabular}{rlrrrrrrlrr}
  \hline
 & Condition & $\Psi$ & Variance & SE & Lower CI & Upper CI & P-value & Fold & N & Delta \\ 
  \hline
1 & A5 & -3.94 & 0.70 & 0.83 & -5.57 & -2.30 & 0.00 & 1 & 167 & 1.00 \\ 
2 & A5 & -3.27 & 0.87 & 0.93 & -5.10 & -1.45 & 0.00 & 2 & 167 & 1.00 \\ 
3 & A5 & -3.50 & 0.61 & 0.78 & -5.04 & -1.97 & 0.00 & 3 & 166 & 1.00 \\ 
4 & Rank 1 Neg & -3.67 & 0.25 & 0.50 & -4.65 & -2.70 & 0.00 & Pooled TMLE & 500 & 1.00 \\ 
   \hline
\end{tabular}
\caption{Rank 1 Negative Marginal Results for NIEHS Data}
\label{tab:inverse_relationships}
\end{table}

\textbf{Table \ref{tab:inverse_relationships}} presents the rank 1 inverse marginal results for the NIEHS data. The first three rows show fold-specific results, with $\Psi$ indicating the expected change in the outcome under the exposure shift compared to the average observed outcome. Here, $A5$ is identified as the top negatively associated mixture variable across all folds. The pooled TMLE column provides the aggregate results, demonstrating that a 1-unit increase in $A5$ leads to an average decrease of -3.67 in endocrine disruption, with a standard error of 0.50 and a 95\% confidence interval of -4.65 to -2.70. These results highlight the consistent negative association of $A5$ with the outcome across different data folds.

\begin{table}[ht]
\hspace{-1cm}\begin{tabular}{rrrrrrrrrrrl}
  \hline
 & Rank & $\Psi$ & Variance & SE & Lower CI & Upper CI & P-value & Fold & N & Delta 1 & Delta 2 \\ 
  \hline
1 & 1 & 19.94 & 0.44 & 0.66 & 18.64 & 21.25 & 0.00 & 1 & 167 & 1.00 & A1 \\ 
2 & 1 & 2.91 & 0.84 & 0.92 & 1.11 & 4.71 & 0.00 & 1 & 167 & 1.00 & A7 \\ 
3 & 1 & 25.74 & 2.46 & 1.57 & 21.9 & 26.2 & 0.00 & 1 & 167 & 1.00 & A1-A7 \\ 
4 & 1 & 2.89 & 2.03 & 1.43 & 1.39 & 5.30 & 0.00 & 1 & 167 & 1.00 & Interaction \\ 
5 & 1 & -2.81 & 1.33 & 1.16 & -5.07 & -0.54 & 0.01 & 2 & 167 & 1.00 & A1 \\ 
6 & 1 & 3.23 & 0.80 & 0.89 & 1.48 & 4.98 & 0.00 & 2 & 167 & 1.00 & A7 \\ 
7 & 1 & 24.35 & 0.62 & 0.78 & 22.81 & 25.89 & 0.00 & 2 & 167 & 1.00 & A1-A7 \\ 
8 & 1 & 23.93 & 5.03 & 2.24 & 19.53 & 28.32 & 0.00 & 2 & 167 & 1.00 & Interaction \\ 
9 & 1 & 3.11 & 0.77 & 0.88 & 1.39 & 4.83 & 0.00 & 3 & 166 & 1.00 & A2 \\ 
10 & 1 & 3.72 & 0.44 & 0.66 & 2.42 & 5.02 & 0.00 & 3 & 166 & 1.00 & A7 \\ 
11 & 1 & 14.46 & 4.50 & 2.12 & 10.30 & 18.62 & 0.00 & 3 & 166 & 1.00 & A2-A7 \\ 
12 & 1 & 7.62 & 3.23 & 1.80 & 4.10 & 11.15 & 0.00 & 3 & 166 & 1.00 & Interaction \\ 
  \hline
\end{tabular}
\caption{Rank 1 Synergistic Interaction Results for NIEHS Data}
\label{tab:synergistic_interactions}
\end{table}

\textbf{Table \ref{tab:synergistic_interactions}} presents the rank 1 synergistic interaction results for the NIEHS data. The table shows fold-specific results, with $\Psi$ indicating the expected change in the outcome under a 1-unit shift in the specified exposure(s). Here, we observe that $A1$ and $A7$ are identified as the top synergistic relationship in two folds, while $A2$ and $A7$ are identified in the remaining fold. The joint shift for $A1$ and $A7$ and for $A2$ and $A7$ reveals significant synergistic interactions built into the data-generating process. The interaction effects reflect the combined effect of shifting both exposures compared to the sum of their individual shifts, indicating strong synergy where the Psi under interaction is positive. 

\begin{table}[ht]
\hspace{-2cm}\begin{tabular}{rlrrrrrrlrrl}
  \hline
 & Rank & $\Psi$ & Variance & SE & Lower CI & Upper CI & P-value & Fold & N & Delta 1 & Delta 2 \\ 
  \hline
1 & Rank 1 & 6.68 & 0.20 & 0.45 & 5.81 & 7.56 & 0.00 & Pooled TMLE & 500 & 1.00 & Var 1 \\ 
2 & Rank 1 & 3.27 & 0.26 & 0.51 & 2.27 & 4.27 & 0.00 & Pooled TMLE & 500 & 1.00 & Var 2 \\ 
3 & Rank 1 & 21.04 & 0.32 & 0.57 & 19.92 & 22.15 & 0.00 & Pooled TMLE & 500 & 1.00 & Joint \\ 
4 & Rank 1 & 11.08 & 1.01 & 1.00 & 9.11 & 13.05 & 0.00 & Pooled TMLE & 500 & 1.00 & Interaction \\ 
  \hline
\end{tabular}
\caption{Pooled TMLE Estimates for Rank 1 Synergy in NIEHS Data}
\label{tab:pooled_synergy}
\end{table}

\textbf{Table \ref{tab:pooled_synergy}} presents the pooled TMLE estimates for the rank 1 synergistic relationships in the NIEHS data.  The first two rows show the average individual effects of the variables involved in the top synergistic relationships, identified as Var 1 and Var 2, here because Var 1 changed across the folds it is the average across $A1$ and $A2$ and Var 2 corresponds to $A7$. The third row represents the joint effect of shifting both exposures simultaneously, while the fourth row shows the average synergistic interaction effect, which is significantly greater than the sum of individual shifts. This pooled estimate reflects the average effect across the identified top synergistic relationships. 

\begin{table}[ht]
\hspace{-1cm}\begin{tabular}{rrrrrrrrrrrl}
  \hline
 & Rank & $\Psi$ & Variance & SE & Lower CI & Upper CI & P-value & Fold & N & Delta  & Type \\ 
  \hline
1 & 1 & -3.97 & 0.71 & 0.84 & -5.62 & -2.32 & 0.00 & 1 & 167 & 1.00 & X5 \\ 
2 & 1 & 3.00 & 0.84 & 0.91 & 1.21 & 4.79 & 0.00 & 1 & 167 & 1.00 & X7 \\ 
3 & 1 & -1.76 & 1.00 & 1.00 & -3.72 & 0.20 & 0.08 & 1 & 167 & 1.00 & X5-X7 \\ 
4 & 1 & -0.78 & 0.93 & 0.96 & -2.67 & 1.11 & 0.42 & 1 & 167 & 1.00 & Interaction \\ 
5 & 1 & -3.48 & 0.74 & 0.86 & -5.16 & -1.79 & 0.00 & 2 & 167 & 1.00 & X5 \\ 
6 & 1 & 3.21 & 0.80 & 0.89 & 1.46 & 4.96 & 0.00 & 2 & 167 & 1.00 & X7 \\ 
7 & 1 & -0.61 & 1.22 & 1.10 & -2.77 & 1.56 & 0.56 & 2 & 167 & 1.00 & X5-X7 \\ 
8 & 1 & -0.34 & 1.08 & 1.04 & -2.38 & 1.70 & 0.74 & 2 & 167 & 1.00 & Interaction \\ 
9 & 1 & -3.45 & 0.54 & 0.73 & -4.89 & -2.01 & 0.00 & 3 & 166 & 1.00 & X5 \\ 
10 & 1 & 3.83 & 0.43 & 0.66 & 2.54 & 5.13 & 0.00 & 3 & 166 & 1.00 & X7 \\ 
11 & 1 & -0.69 & 1.17 & 1.08 & -2.81 & 1.43 & 0.50 & 3 & 166 & 1.00 & X5-X7 \\ 
12 & 1 & -1.07 & 0.90 & 0.95 & -2.94 & 0.79 & 0.27 & 3 & 166 & 1.00 & Interaction \\ 
  \hline
\end{tabular}
\caption{K-Fold Specific Antagonistic Rank 1 Results for NIEHS Data}
\label{tab:kfold_antagonism}
\end{table}

\textbf{Table \ref{tab:kfold_antagonism}} presents the k-fold specific antagonistic rank 1 results for the NIEHS data. The $\Psi$ column represents the expected change in the outcome under a 1-unit shift in the specified exposures. For the first fold, variable A5 shows a negative shift of -3.97, indicating a strong inverse association. Similarly, variable A7 shows a positive shift of 3.00. The combined shift of A5 and A7 results in an estimated change of -1.76, while the interaction term, which compares the joint shift to the two marginal shifts, shows a negative shift of -0.78. These patterns are consistent across all three folds, demonstrating significant evidence for antagonistic interactions between these variables.

\subsubsection{Comparison to Existing Methods}
Quantile g-computation \cite{Keil2019}, prevalent in environmental epidemiology for mixture analysis, estimates the effects of uniformly increasing exposures by one quantile, based on linear model assumptions. This method quantizes mixture components, summing the linear model's coefficients to form a summary measure ($\Psi$) for joint impact assessment. However, it inherently assumes additive, monotonic exposure-response relationships, overlooking complex, potentially nonlinear interactions typical in mixtures, like in endocrine disrupting compounds. Consequently, this method might not accurately capture the nuanced dynamics of mixed exposures, especially when interactions vary with other variable levels. 

We run quantile g-computation on the NIEHS data using 4 quantiles with no interactions to investigate results using this model. The size of the scaled effect (positive direction, sum of positive coefficients) was 6.28 and included $A_1, A_2, A_3, A_7$ and the scaled effect size (negative direction, sum of negative coefficients) was -3.68 and included $A_4, A_5, A_6$. Compared to NIEHS ground truth, $A_3, A_6$ are incorrectly included in these estimates. However, the positive and negative associations for the other variables are correct. Next, because we expect interactions to exist in the mixture data, we would like to assess for them but the question is which interaction terms to include? Our best guess is to include interaction terms for all exposures. We do this and show results in \textbf{Table \ref{tab:q_comp_itxns}}.

\begin{table}[ht]
\centering
\begin{tabular}{rrrrrr}
  \toprule
 & Estimate & Std. Error & Lower CI & Upper CI & Pr($>$$|$t$|$) \\ 
  \midrule
(Intercept) & 21.29 & 1.58 & 18.19 & 24.39 & 0.00 \\ 
  psi1 & 0.02 & 1.62 & -3.16 & 3.20 & 0.99 \\ 
  psi2 & 0.59 & 0.67 & -0.71 & 1.90 & 0.37 \\ 
   \bottomrule
\end{tabular}
\caption{Quantile G-Computation Interaction Results from NIEHS Synthetic Data}
\label{tab:q_comp_itxns}
\end{table}

In \textbf{Table \ref{tab:q_comp_itxns}} $\Psi_1$ is the summary measure for the main effects and $\Psi_2$ for interactions. As can be seen, when including all interactions, neither of the estimates are significant. Of course, this is to be expected given the number of parameters in the model and the sample size $n=500$. However, moving forward with interaction assessment is difficult; if we were to assess for all 2-way interaction of 7 exposures, the number of sets is 21 and with 3-way interactions is 35. We would have to run these many models and then correct for multiple testing. Hopefully, this example shows why mixtures are inherently a data-adaptive problem and why popular methods such as this, although succinct and interpretable, fall short even in a simple synthetic data set. 

\subsection{Methods Applied to NHANES Dataset}

The National Health and Nutrition Examination Survey (NHANES) 2001-2002 cycle serves as the foundation of our analysis. This data source, known for its credibility in the public health domain, boasts interviews with 11,039 individuals. Of this subset, 4,260 provided blood samples and willingly consented to DNA analysis. The data set used aligns with that of Mitro et al. \cite{mitro2016cross} focusing on the correlation between persistent exposure to organic pollutants (POPs), specifically those binding to the aryl hydrocarbon receptor (AhR) and extended leukocyte telomere length (LTL). However, this subset was further refined to ensure complete exposure data, yielding 1007 participants for our analysis, compared to 1003 in the initial study by Mitro et al. In alignment with protocols detailed by Mitro et al. \cite{mitro2016cross}, exposure was quantified, focusing on 18 congeners. These include 8 non-dioxin-like PBCs, 2 non-ortho PCBs, 1 mono-ortho PCB, 4 Dioxins, and 4 Furans. All congeners underwent lipid serum adjustments using an enzymatic summation methodology. The telomere length, was analyzed using the quantitative polymerase chain reaction (qPCR) methodology \cite{mitro2016cross}. This technique measures the T/S ratio by comparing the length of the telomere to a standardized reference DNA. To enhance accuracy, samples underwent triple assays in duplicate, generating an average from six data points. The CDC conducted a blinded quality control assessment. Our modeling accounted for several covariates, including demographic factors such as age, sex, race / ethnicity and level of education, as well as health indicators such as BMI, serum cotinine, and blood cell distribution and count. Categorizations for race/ethnicity, education, and BMI are consistent with previous studies. Furthermore, this comprehensive dataset is integrated into the \texttt{InterXshift} package for replicable analysis. 

\begin{figure*}
  \centerline{\includegraphics[width=7.0in]{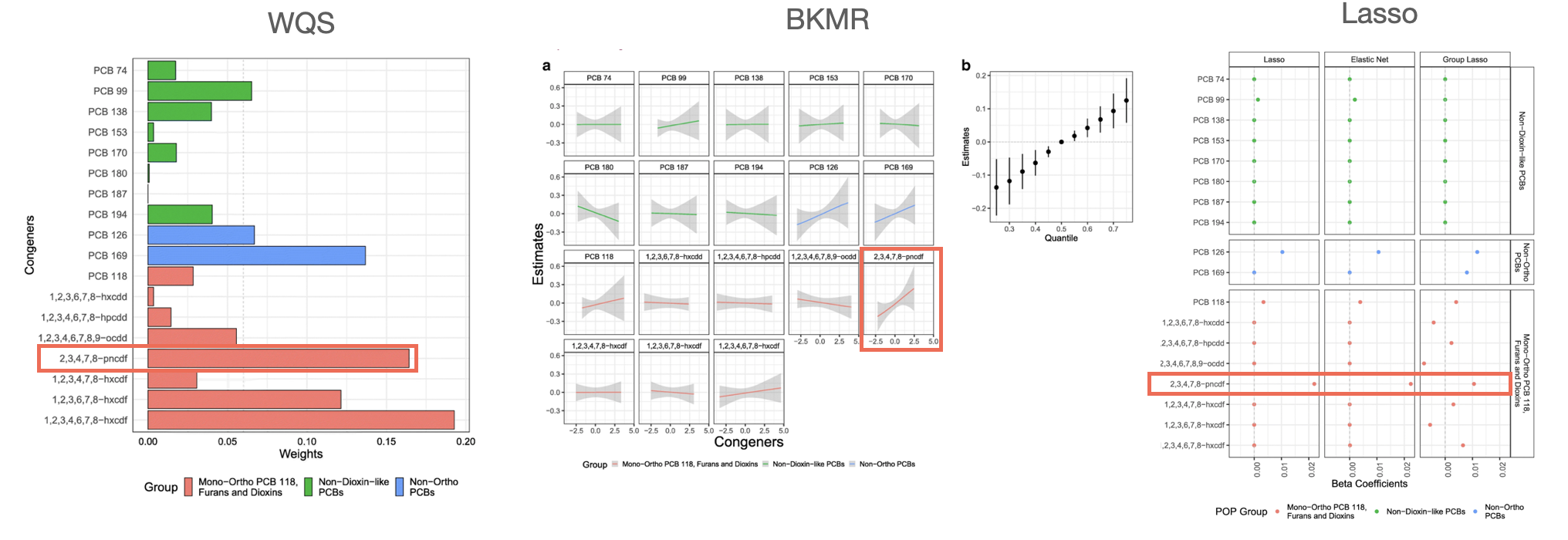}}
  \caption{Mixture Methods Results from Gibson et. al.}
\label{fig:gibson}
\end{figure*}

\cite{gibson2019overview} reanalyzed this same data using more contemporary statistical methods. They found that clustering methods identified high, medium, and low POP exposure groups with longer log-LTL observed in the high exposure group. Principal component analysis (PCA) and exploratory factor analysis (EFA) revealed positive associations between overall POP exposure and specific POPs with log-LTL. Penalized regression methods identified three congeners (PCB 126, PCB 118, and furan 2,3,4,7,8-pncdf) as potentially toxic agents. WQS identified six POPs (furans 1,2,3,4,6,7,8-hxcdf, 2,3,4,7,8-pncdf, and 1,2,3,6,7,8-hxcdf, and PCBs 99, 126, 169) as potentially toxic agents with a positive overall effect of the POP mixture. BKMR found a positive linear association with furan 2,3,4,7,8-pncdf, suggestive evidence of linear associations with PCBs 126 and 169, a positive overall effect of the mixture, but no interactions among congeners. These results (in the supervised methods) controlled for the same covariates. \textbf{Figure \ref{fig:gibson}} shows the results for WQS, BKMR and Lasso regression from their paper. We highlight furan 2,3,4,7,8-pncdf as the chemical which shows consistent positive association with LTL across the methods. As such, we might expect that this chemical would consistently be found as the top positive marginal impact based on our target parameter across all the folds. 

We configured our methodology using default learners and employed a 5-fold CV. Given the data range, a $\delta$ value of half a standard deviation was used for all exposures, denoting a focus on counterfactual changes in telomere length for an increase in half a standard deviation across exposures that predict the telomere length depending on the scale of exposure.

\subsection{NHANES Furan Results}

Generally, when applying our InterXshift method to identify the top positive and negative marginal effects; and the top synergistic and antagonistic interactions, we do not find consistent estimates for these parameters. \textbf{Table \ref{tab:NHANES_rank1_pos_results}} shows the estimates of the top positive association between the folds. Although we see 2,3,4,7,8-pncdf is identified as the top positive effect in 3 of the folds, 2 of the folds estimate a different chemical as the top positive effect; therefore, the estimates are inconsistent. Likewise, the estimates for the top inverse association with LTL were also inconsistent across the folds. The top synergistic and antagonistic interactions were also inconsistent and, therefore, indicate no interaction in the mixed POP exposure. 

\begin{table}[ht]
\hspace{-2cm}\begin{tabular}{rlrrrrrrlrr}
  \hline
 & Condition & Psi & Variance & SE & Lower CI & Upper CI & P-value & Fold & N & Delta \\ 
  \hline
1 & 2,3,4,7,8-pncdf & 0.02 & 0.00 & 0.02 & -0.02 & 0.05 & 0.42 & 1 & 202 & 2.88 \\ 
2 & 2,3,4,7,8-pncdf & 0.01 & 0.00 & 0.02 & -0.03 & 0.04 & 0.64 & 2 & 202 & 2.88 \\ 
3 & PCB74 & -0.05 & 0.00 & 0.03 & -0.11 & 0.02 & 0.18 & 3 & 201 & 6794.89 \\ 
4 & 1,2,3,4,7,8-hxcdf & 0.00 & 0.00 & 0.02 & -0.04 & 0.05 & 0.92 & 4 & 201 & 2.49 \\ 
5 & 2,3,4,7,8-pncdf & 0.03 & 0.00 & 0.02 & -0.01 & 0.08 & 0.12 & 5 & 201 & 2.88 \\ 
6 & Rank 1 & -0.00 & 0.00 & 0.01 & -0.03 & 0.02 & 0.74 & Pooled TMLE & 1007 & 1361.20 \\ 
   \hline
\end{tabular}
\caption{Rank 1 Results for Different Conditions}
\label{tab:NHANES_rank1_pos_results}
\end{table}

\section{Discussion}

In our study, we developed a semiparametric statistical model to discover and estimate interaction effects in mixed exposures. Using data-adaptive target parameters paired with TMLE within a cross-validation framework, we achieved asymptotically unbiased estimates of identifying and estimating the top interaction effects in a mixed exposure. Our approach offers several advantages, including robustness to the number of exposures and the complexity of the data-generating process, and provides interpretable results that improve understanding of drug synergies and environmental health impacts. Unlike traditional interaction analysis methods, our model minimizes assumptions, yielding more reliable and interpretable estimates of interaction effects, and focuses on causal quantities. Our methodology faces the primary limitation that the identification of exposure sets that are estimated for the top marginal, synergistic, and antagonistic interaction may vary between cross-validation folds, introducing variability that demands careful and consistent reporting to ensure reproducibility and reliability of results. However, rather than a complete limitation, this indicates the strength of the association in subsets of the data. If the ``signal`` is consistent, then the findings for each parameter will be consistent. 

In future research, our aim is to improve the estimation of density estimation ratios and to evaluate the variability of estimates using various methods for estimation of density ratios. We can also focus beyond two-way interactions to include higher-order complexities. We also intend to investigate alternative TMLE strategies that may offer better performance, especially in smaller samples. To facilitate adoption, we have developed the \texttt{InterXshift} R package, available on GitHub, offering researchers software to apply our proposed methodology for the analysis of mixtures, this package has been peer reviewed by the Journal of Open-Source Software \cite{McCoy2023}. 

\bibliography{bibliography}

\begin{thebibliography}{30}
\providecommand{\natexlab}[1]{#1}
\providecommand{\url}[1]{\texttt{#1}}
\expandafter\ifx\csname urlstyle\endcsname\relax
  \providecommand{\doi}[1]{doi: #1}\else
  \providecommand{\doi}{doi: \begingroup \urlstyle{rm}\Url}\fi

\bibitem[Bobb et~al.(2014)Bobb, Valeri, {Claus Henn}, Christiani, Wright, Mazumdar, Godleski, and Coull]{Bobb2014}
Jennifer~F. Bobb, Linda Valeri, Birgit {Claus Henn}, David~C. Christiani, Robert~O. Wright, Maitreyi Mazumdar, John~J. Godleski, and Brent~A. Coull.
\newblock {Bayesian kernel machine regression for estimating the health effects of multi-pollutant mixtures}.
\newblock \emph{Biostatistics}, 16\penalty0 (3):\penalty0 493--508, 2014.
\newblock ISSN 14684357.
\newblock \doi{10.1093/biostatistics/kxu058}.

\bibitem[Chen and Guestrin(2016)]{xgboost}
Tianqi Chen and Carlos Guestrin.
\newblock {XGBoost}: A scalable tree boosting system.
\newblock In \emph{Proceedings of the 22nd ACM SIGKDD International Conference on Knowledge Discovery and Data Mining}, KDD '16, pages 785--794, New York, NY, USA, 2016. ACM.
\newblock ISBN 978-1-4503-4232-2.
\newblock \doi{10.1145/2939672.2939785}.
\newblock URL \url{http://doi.acm.org/10.1145/2939672.2939785}.

\bibitem[Daniel et~al.(2013)Daniel, Cousens, De~Stavola, Kenward, and Sterne]{Daniel2013}
Rhian~M Daniel, Simon~N Cousens, Bianca~L De~Stavola, Michael~G Kenward, and Jonathan~AC Sterne.
\newblock Methods for dealing with time-dependent confounding.
\newblock \emph{Statistics in Medicine}, 32\penalty0 (9):\penalty0 1584--1618, 2013.
\newblock \doi{10.1002/sim.5686}.

\bibitem[D{\'{i}}az et~al.(2023)D{\'{i}}az, Williams, Hoffman, and Schenck]{Diaz2023}
Iv{\'{a}}n D{\'{i}}az, Nicholas Williams, Katherine~L. Hoffman, and Edward~J. Schenck.
\newblock {Nonparametric Causal Effects Based on Longitudinal Modified Treatment Policies}.
\newblock \emph{Journal of the American Statistical Association}, 118\penalty0 (542):\penalty0 846--857, 2023.
\newblock ISSN 1537274X.
\newblock \doi{10.1080/01621459.2021.1955691}.

\bibitem[Garc{\'{i}}a-Villarino et~al.(2022)Garc{\'{i}}a-Villarino, Signes-Pastor, Karagas, Ria{\~{n}}o-Gal{\'{a}}n, Rodr{\'{i}}guez-Dehli, Grimalt, Junqu{\'{e}}, Fern{\'{a}}ndez-Somoano, and Tard{\'{o}}n]{Garcia-Villarino2022}
Miguel Garc{\'{i}}a-Villarino, Antonio~J. Signes-Pastor, Margaret~R. Karagas, Isolina Ria{\~{n}}o-Gal{\'{a}}n, Cristina Rodr{\'{i}}guez-Dehli, Joan~O. Grimalt, Eva Junqu{\'{e}}, Ana Fern{\'{a}}ndez-Somoano, and Adonina Tard{\'{o}}n.
\newblock {Exposure to metal mixture and growth indicators at 4–5 years. A study in the INMA-Asturias cohort}.
\newblock \emph{Environmental Research}, 204, 2022.
\newblock ISSN 10960953.
\newblock \doi{10.1016/j.envres.2021.112375}.

\bibitem[Gibson et~al.(2019)Gibson, Nunez, Abuawad, Zota, Renzetti, Devick, Gennings, Goldsmith, Coull, and Kioumourtzoglou]{gibson2019overview}
Elizabeth~A Gibson, Yanelli Nunez, Ahlam Abuawad, Ami~R Zota, Stefano Renzetti, Katrina~L Devick, Chris Gennings, Jeff Goldsmith, Brent~A Coull, and Marianthi-Anna Kioumourtzoglou.
\newblock An overview of methods to address distinct research questions on environmental mixtures: an application to persistent organic pollutants and leukocyte telomere length.
\newblock \emph{Environmental Health}, 18:\penalty0 1--16, 2019.

\bibitem[Hubbard et~al.(2018)Hubbard, Kennedy, and Laan]{hubbard_book_da}
Alan Hubbard, Chris Kennedy, and Mark Laan.
\newblock \emph{Data-Adaptive Target Parameters}, pages 125--142.
\newblock 01 2018.
\newblock ISBN 978-3-319-65303-7.
\newblock \doi{10.1007/978-3-319-65304-4_9}.

\bibitem[Hubbard et~al.(2016)Hubbard, Kherad-Pajouh, and {Van Der Laan}]{Hubbard2016}
Alan~E. Hubbard, Sara Kherad-Pajouh, and Mark~J. {Van Der Laan}.
\newblock {Statistical Inference for Data Adaptive Target Parameters}.
\newblock \emph{International Journal of Biostatistics}, 12\penalty0 (1):\penalty0 3--19, 2016.
\newblock ISSN 15574679.
\newblock \doi{10.1515/ijb-2015-0013}.

\bibitem[{Iv{\'{a}}n D{\'{i}}az Mu{\~{n}}oz and Mark van der Laan*}(2012)]{diaz_2012}
{Iv{\'{a}}n D{\'{i}}az Mu{\~{n}}oz and Mark van der Laan*}.
\newblock {Population Intervention Causal Effects Based on Stochastic Interventions}.
\newblock \emph{Biometrics.}, 68\penalty0 (2):\penalty0 541--549, 2012.
\newblock ISSN 15378276.
\newblock \doi{10.1111/j.1541-0420.2011.01685.x.Population}.
\newblock URL \url{https://www.ncbi.nlm.nih.gov/pmc/articles/PMC3624763/pdf/nihms412728.pdf}.

\bibitem[Keil et~al.(2019)Keil, Buckley, O'Brien, Ferguson, Zhao, and White]{Keil2019}
Alexander~P. Keil, Jessie~P. Buckley, Katie~M. O'Brien, Kelly~K. Ferguson, Shanshan Zhao, and Alexandra~J. White.
\newblock {A quantile-based g-computation approach to addressing the effects of exposure mixtures}.
\newblock \emph{arXiv}, 128\penalty0 (April):\penalty0 1--10, 2019.
\newblock ISSN 23318422.
\newblock \doi{10.1097/01.ee9.0000606120.58494.9d}.

\bibitem[Lengerich et~al.(2020)Lengerich, Tan, Chang, Hooker, and Caruana]{Lengerich2020}
Benjamin Lengerich, Sarah Tan, Chun~Hao Chang, Giles Hooker, and Rich Caruana.
\newblock {Purifying Interaction Effects with the Functional ANOVA: An Efficient Algorithm for Recovering Identifiable Additive Models}.
\newblock \emph{Proceedings of Machine Learning Research}, 108:\penalty0 2402--2412, 2020.
\newblock ISSN 26403498.

\bibitem[Li et~al.(2022)Li, Rosete, Coyle, Phillips, Hejazi, Malenica, Arnold, Benjamin-Chung, Mertens, Colford~Jr, van~der Laan, and Hubbard]{best_tmle}
Haodong Li, Sonali Rosete, Jeremy Coyle, Rachael~V. Phillips, Nima~S. Hejazi, Ivana Malenica, Benjamin~F. Arnold, Jade Benjamin-Chung, Andrew Mertens, John~M. Colford~Jr, Mark~J. van~der Laan, and Alan~E. Hubbard.
\newblock Evaluating the robustness of targeted maximum likelihood estimators via realistic simulations in nutrition intervention trials.
\newblock \emph{Statistics in Medicine}, 41\penalty0 (12):\penalty0 2132--2165, 2022.
\newblock \doi{https://doi.org/10.1002/sim.9348}.
\newblock URL \url{https://onlinelibrary.wiley.com/doi/abs/10.1002/sim.9348}.

\bibitem[McCoy et~al.(2023)McCoy, Schuler, Hubbard, and van~der Laan]{McCoy2023}
David McCoy, Alejandro Schuler, Alan Hubbard, and Mark van~der Laan.
\newblock Supernova: Semi-parametric identification and estimation of interaction and effect modification in mixed exposures using stochastic interventions in r.
\newblock \emph{Journal of Open Source Software}, 8\penalty0 (91):\penalty0 5422, 2023.
\newblock \doi{10.21105/joss.05422}.
\newblock URL \url{https://doi.org/10.21105/joss.05422}.

\bibitem[McEligot et~al.(2020)McEligot, Poynor, Sharma, and Panangadan]{McEligot2020}
Archana~J. McEligot, Valerie Poynor, Rishabh Sharma, and Anand Panangadan.
\newblock {Logistic lasso regression for dietary intakes and breast cancer}.
\newblock \emph{Nutrients}, 12\penalty0 (9):\penalty0 1--14, 2020.
\newblock ISSN 20726643.
\newblock \doi{10.3390/nu12092652}.

\bibitem[Mitro et~al.(2016)Mitro, Birnbaum, Needham, and Zota]{mitro2016cross}
Susanna~D Mitro, Linda~S Birnbaum, Belinda~L Needham, and Ami~R Zota.
\newblock Cross-sectional associations between exposure to persistent organic pollutants and leukocyte telomere length among us adults in nhanes, 2001--2002.
\newblock \emph{Environmental health perspectives}, 124\penalty0 (5):\penalty0 651--658, 2016.

\bibitem[Narisetty et~al.(2019)Narisetty, Mukherjee, Chen, Gonzalez, and Meeker]{Narisetty2019Selection}
Naveen~N. Narisetty, Bhramar Mukherjee, Yi-Hau Chen, Richard Gonzalez, and John~D. Meeker.
\newblock Selection of nonlinear interactions by a forward stepwise algorithm: Application to identifying environmental chemical mixtures affecting health outcomes.
\newblock \emph{Statistical Medicine}, 38\penalty0 (9):\penalty0 1582--1600, Apr 2019.
\newblock \doi{10.1002/sim.8059}.

\bibitem[Roberts and Martin(2006)]{Roberts2006}
Steven Roberts and Michael~A. Martin.
\newblock {Using supervised principal components analysis to assess multiple pollutant effects}.
\newblock \emph{Environmental Health Perspectives}, 114\penalty0 (12):\penalty0 1877--1882, 2006.
\newblock ISSN 00916765.
\newblock \doi{10.1289/ehp.9226}.

\bibitem[Robins(1987)]{Robins1987}
J.~M. Robins.
\newblock {Addendum To 'a New Approach To Causal Inference in Mortality Studies With a Sustained Exposure Period - Application To Control of the Healthy Worker Survivor Effect'.}
\newblock \emph{Computers \& mathematics with applications}, 14\penalty0 (9-12):\penalty0 923--945, 1987.
\newblock ISSN 00974943.
\newblock \doi{10.1016/0898-1221(87)90238-0}.

\bibitem[Rothman et~al.(1980)Rothman, Greenland, and Walker]{rothman1980concepts}
Kenneth~J Rothman, Sander Greenland, and Alexander~M Walker.
\newblock Concepts of interaction.
\newblock \emph{Am J Epidemiol}, 112\penalty0 (4):\penalty0 467--470, 1980.

\bibitem[Rubin(1974)]{Rubin1974-od}
Donald~B Rubin.
\newblock Estimating causal effects of treatments in randomized and nonrandomized studies.
\newblock \emph{J. Educ. Psychol.}, 66\penalty0 (5):\penalty0 688--701, October 1974.

\bibitem[Rubin(2005)]{Rubin2005-by}
Donald~B Rubin.
\newblock Causal inference using potential outcomes.
\newblock \emph{J. Am. Stat. Assoc.}, 100\penalty0 (469):\penalty0 322--331, March 2005.

\bibitem[Tay et~al.(2023)Tay, Narasimhan, and Hastie]{elasticnet}
J.~Kenneth Tay, Balasubramanian Narasimhan, and Trevor Hastie.
\newblock Elastic net regularization paths for all generalized linear models.
\newblock \emph{Journal of Statistical Software}, 106\penalty0 (1):\penalty0 1--31, 2023.
\newblock \doi{10.18637/jss.v106.i01}.

\bibitem[van~der Laan(2006)]{Van_der_Laan2006-ro}
Mark~J van~der Laan.
\newblock Statistical inference for variable importance.
\newblock \emph{Int. J. Biostat.}, 2\penalty0 (1), February 2006.

\bibitem[van~der Laan~Mark et~al.(2007)van~der Laan~Mark, C, and E.]{SL_2008}
J.~van~der Laan~Mark, Polley~Eric C, and Hubbard~Alan E.
\newblock Super learner.
\newblock \emph{Statistical Applications in Genetics and Molecular Biology}, 6\penalty0 (1):\penalty0 1--23, 2007.
\newblock URL \url{https://EconPapers.repec.org/RePEc:bpj:sagmbi:v:6:y:2007:i:1:n:25}.

\bibitem[VanderWeele(2009)]{vanderweele2009distinction}
Tyler~J VanderWeele.
\newblock On the distinction between interaction and effect modification.
\newblock \emph{Epidemiology}, 20\penalty0 (6):\penalty0 863--871, 2009.

\bibitem[VanderWeele and Knol(2014)]{vanderweele2014tutorial}
Tyler~J VanderWeele and Mirjam~J Knol.
\newblock A tutorial on interaction.
\newblock \emph{Epidemiologic Methods}, 3\penalty0 (1):\penalty0 33--72, 2014.

\bibitem[VanderWeele and Robins(2007)]{vanderweele2007directed}
Tyler~J VanderWeele and James~M Robins.
\newblock Directed acyclic graphs, sufficient causes, and the properties of conditioning on a common effect.
\newblock \emph{Am J Epidemiol}, 166\penalty0 (10):\penalty0 1096--1104, 2007.

\bibitem[Wang et~al.(2018)Wang, Mukherjee, and Park]{Wang2018}
Xin Wang, Bhramar Mukherjee, and Sung~Kyun Park.
\newblock Associations of cumulative exposure to heavy metal mixtures with obesity and its comorbidities among u.s. adults in nhanes 2003-2014.
\newblock \emph{Environment International}, 121\penalty0 (Pt 1):\penalty0 683--694, 2018.
\newblock \doi{10.1016/j.envint.2018.09.035}.

\bibitem[Wright and Ziegler(2017)]{ranger}
Marvin~N. Wright and Andreas Ziegler.
\newblock {ranger}: A fast implementation of random forests for high dimensional data in {C++} and {R}.
\newblock \emph{Journal of Statistical Software}, 77\penalty0 (1):\penalty0 1--17, 2017.
\newblock \doi{10.18637/jss.v077.i01}.

\bibitem[Zheng and van~der Laan(2010)]{Zheng2010}
Wenjing Zheng and MJ~van~der Laan.
\newblock {Asymptotic theory for cross-validated targeted maximum likelihood estimation}.
\newblock \emph{U.C. Berkeley Division of Biostatistics Working Paper Series}, \penalty0 (273), 2010.
\newblock URL \url{http://biostats.bepress.com/ucbbiostat/paper273/}.

\end{thebibliography}

\end{document}